# Evolution of channel flow and Darcy's law beyond the critical Reynolds number

*Xiaohui Deng, Ping Sheng*[*]

*Department of Physics, The Hong Kong University of Science and Technology, Clear Water Bay, Kowloon, Hong Kong, China*

[*]sheng@ust.hk

For incompressible channel flow there is a critical state, characterized by a critical Reynolds number $Re_c$ and a critical wavevector $m_c$ along the channel direction, beyond which the channel flow becomes unstable in the linear regime. In this work we investigate the channel flow beyond the critical state and find the existence of a new fluctuating, quasi-stationary flow with ~15% reduced flow rate from the Poiseuille flow in the laminar regime. Vortices and anti-vortices accompany the time-varying flow profile. Our study is facilitated by the analytical solution of the linearized, incompressible, three dimensional (3D) Navier-Stokes (NS) equation in the channel geometry, with the Navier boundary condition; alternatively denoted as the hydrodynamic modes (HMs). By using the HMs as the complete mathematical basis for expanding the velocity in the NS equation, the $Re_c$ is evaluated to 5-digit accuracy when compared to the well-known Orszag result, without invoking the Orr-Sommerfeld equation. Beyond $Re_c$, we find the fluctuating equilibrium flow profile to be decomposable into the Poiseuille flow component and a counter-flow component that differs from any of the pressure-driven channel flows. In particular, the counter flow is found to comprise multiple HMs, some with opposite flow direction, that can lead to a net boundary reaction force along the counter flow direction. The latter is necessary to satisfy the Newton's law, as a reduced flow rate requires the generation of boundary reaction force for altering the centre of mass momentum. Experimental verification of the predictions is discussed.

## I. Introduction

As a pillar of classical physics, the Navier-Stokes equation governs a broad range of fluid dynamic behaviours [1-13]. For incompressible fluid, the non-dimensionalized NS equation has the following simple form:

$$\frac{\partial \boldsymbol{v}}{\partial t} + (\boldsymbol{v}\cdot\nabla)\boldsymbol{v} = -\nabla p + \frac{1}{\text{Re}}\nabla^2\boldsymbol{v}, \tag{1a}$$

$$\nabla \cdot \mathbf{v} = 0 , \tag{1b}$$

where $\mathbf{v}$ denotes velocity, $p$ the pressure, and Re is the Reynolds number [2]. In Eq. (1a), $(\mathbf{v} \cdot \nabla)\mathbf{v}$ is the nonlinear term that gives rise to much of the complexity and richness of the fluid dynamic behavior.

Consider the incompressible channel flow between two parallel solid boundaries separated by a distance $2H$ along the $z$ direction, driven by a constant pressure gradient along the flow direction. We divide the velocity $\mathbf{v}$ into two components: $\mathbf{v} = \mathbf{u}^P + \mathbf{u}$, where $\mathbf{u}^P$ is the Poiseuille flow velocity under an externally applied pressure, and $\mathbf{u}$ being the perturbation component, both with the Navier slip boundary condition at the fluid-solid interfaces. If we express length in units of $H$, mass density in units of the fluid mass density $\rho$, time in units of $H/U_0$, where $U_0$ denotes the maximum velocity of the Poiseuille flow [2,3] at the symmetry plane $z=0$ under the non-slip boundary condition, then Re=$\rho H U_0 / \eta$, where $\eta$ denotes the shear viscosity. The Navier boundary condition [2,4,14-16] has the form:

$$\text{Velocity components at } z = \pm 1 = \begin{cases} v_{x,y} = [-\text{sgn}(z)] l_s \dfrac{\partial v_{x,y}}{\partial z} , \\ v_z = 0 \end{cases} \tag{2}$$

where $l_s$ is the slip length. The laminar Poiseuille profile is given by

$$u_x^P(z) = \left(1 - z^2 + 2l_s\right), \quad u_{y,z}^P = 0. \tag{3a}$$

The associated flow rate $Q^P$ per unit area in the $yz$ plane and the associated pressure gradient are given by

$$Q^P = 4\left(\frac{1}{3} + l_s\right), \quad \frac{\partial P_0}{\partial x} = -\frac{2}{\text{Re}}, \quad \frac{\partial P_0}{\partial y} = \frac{\partial P_0}{\partial z} = 0 , \tag{3b}$$

where $P_0$ denotes the applied pressure that gives rise to the Poiseuille flow. Equations (3a, 3b) represent the Darcy's law [5-7] in the simplest channel geometry. It expresses a steady channel flow resulting from the externally applied pressure drop, force-balanced by the dissipative viscous force inherent in the flow profile. It is noted that $(\mathbf{u}^P \cdot \nabla)\mathbf{u}^P = 0$, hence theoretically the Poiseuille flow can persist to any Reynolds number. However, this cannot be the case because it has been shown that the laminar flow becomes unstable and can blow up with an infinitesimal perturbation, such as those caused by thermal fluctuations, beyond a critical Reynolds number Re$_c$ [1,2,8-10]. Since physically the channel flow is constrained by a finite energy input, hence what happens beyond Re$_c$ is an intriguing question that is the focus of this study.

By formulating the linearized NS equation into an eigenvalue problem as shown below, we have obtained the analytical solution of the complete set of eigenfunctions for the 3D channel flow with the Navier boundary condition. By expanding $\boldsymbol{u}$ in terms of the HMs [17-20], with $\boldsymbol{u}^P$ being the zeroth order term (input) in the expansion, the time-dependent expansion coefficients become the variables to be solved. That is, Eq. (1a), with the constraint of Eq. (1b), is reduced to an infinitely coupled autonomous system of ODEs for the expansion coefficients, in which time $t$ is the only independent variable, with Re and $l_s$ being the two input parameters.

The use of analytical HMs as the basis function has the advantage of satisfying both the Navier boundary condition and the incompressibility condition, in contrast to the traditional basis functions such as the Fourier basis or the finite-element basis. Hence from the numerical point of view a small number of HMs can already attain high accuracy in the evaluation of the critical Reynolds number, without invoking the Orr-Sommerfeld (OR) equation [4,8]. This point is shown below.

In the system of coupled ODEs, the nonlinear term, $(\boldsymbol{u}\cdot\nabla)\boldsymbol{u}$, takes the form of a third-rank tensor that couples pairs of the expansion coefficients to effect the time variation on the third. Without the nonlinear term, the right hand side of the equation represents a matrix acting on the coefficient vector. By solving for the eigenvalues of the matrix for a finite set of HMs, we found the critical Re$_c$, accompanied by a critical wavevector $m_c$, as the lowest Re value at which the real part of one of the eigenvalues first changes sign, signaling the onset of instability. Our Re$_c$ is accurate to five significant digits when compared with the standard result of Orszag [8] for the non-slip boundary condition, without invoking the OR equation. The effect of a finite slip length on Re$_c$ was also obtained and shown to agree with the prior results [21] on its variation with increasing $l_s$.

Since each HM represents an independent degree of freedom, in a thermal bath each can be excited with $k_B T/2$ of energy, where $k_B$ denotes Boltzmann's constant and $T$ is temperature [22-24]. We examined the Poiseuille flow attendant with the excitation of the HMs by time-evolving the coupled autonomous set of ODEs. It was found that below Re$_c$ the excited HMs decay rapidly [2,8,10], and the Poiseuille flow is stable. However, above Re$_c$ the picture changes dramatically. Under the problem setup where we vary Re by changing $\eta$ while simultaneously keeping $U_0$ constant (by stipulating the applied pressure gradient to vary as 1/Re), it is found that as we numerically time-evolve the autonomous

ODE set, the net flow rate initially decreases but eventually settles at a fluctuating equilibrium state that is statistically stationary.

In what follows, we first present the analytical solution of the HMs for the 3D channel flow, in conjunction with a discussion on their properties. This is followed by the evaluation of the critical state and its parameter values, as well as the effect of a finite slip length. By expanding the perturbation velocity field $\boldsymbol{u}$ in terms of the HMs, we reduce linearized NS equation to a system of autonomous ODEs for the expansion coefficients in which $t$ is the only independent variable. Results of the numerical time evolution are presented, with the initial states statistically sampled from a Gaussian distribution for simulating thermal excitations. A new fluctuating equilibrium state is revealed that exhibits a reduced net flow rate. In particular, interesting features of the flow profiles in the equilibrium state are presented through the lens of force balance. We give a brief discussion on the potential experimental verification of the predictions at the end of the manuscript, and conclude with a brief recapitulation. Details on the derivation of the analytical solution are given in the Supplementary Materials (SM), so as not to detract from the main line of exposition.

## II. Formulation of the eigenvalue problem and its analytical solution

The HMs are the eigenfunctions of the linearized version of the incompressible NS equation [17,18] with the Navier boundary condition at the solid/liquid interface. They are given by [2,17,18,25]:

$$-\lambda \boldsymbol{u} = -\nabla p + \frac{1}{\text{Re}} \nabla^2 \boldsymbol{u}; \qquad \nabla \cdot \boldsymbol{u} = 0 , \tag{4a}$$

$$u_n|_{z=\pm 1} = 0, \quad \left( l_s \frac{\partial u_\tau}{\partial z} \pm u_\tau \right)_{z=\pm 1} = 0 , \tag{4b}$$

where $u_n$ is the component of the velocity field normal to the boundary, and $u_\tau$ is its tangential component. In Eq. (4) we have used the condition that the time variation of $\boldsymbol{u}$ is given by $e^{-\lambda t}$, i.e., the eigenvalue denotes the exponential decay rate of the eigenfunction. The HMs are transient in nature; they cannot be sustained without the input of external energy. Without the loss of generality, periodic boundary condition is imposed on $\boldsymbol{u}$ and $p$ in the $xy$ plane. Depending on the spatial dependence of the eigenfunctions, we classify the HMs into 3 disjoint groups: 1D HMs, 2D HMs and 3D HMs. In each group, according to their spatial symmetry property of the velocity components $u_x, u_y$ with respect to the $z = 0$ plane, they are further classified into the symmetric HMs group and the antisymmetric HMs group. Among these, the 2D and 1D HMs can be further grouped into two different branches, for flow in either the $xz$ plane (denoted the $x$ branch) or the $yz$ plane (denoted the $y$ branch). The 1D and 2D HMs have already been introduced in previous publications [17,18]. Below we focus on the 3D HMs. We emphasize that compared to the NS equation constrained in the 2D spatial domain where vortex stretching is absent and only diffusive exchange of angular momentum exists, in 3D the vortex stretching plays an important role [26,27]. Hence the 3D HMs are crucial for evaluating all possible manifestations of the velocity and vortices patterns.

For the 3D HMs, $\boldsymbol{u}$ is periodic along both the $x, y$ directions, characterized by wavevectors $m$ and $k$, respectively. Here $m \neq 0$, $k \neq 0$ are *real, continuous* variables, implying infinite periodicity in the absence of a length scale defined either by the computational domain, or by a naturally-occurring length scale. In anticipation of later developments, we note that the emergence of a critical Reynolds number is accompanied by the appearance of a length scale in the $xy$ plane, which can be used as the inverse unit for the discretization of $m$, $k$ in the numerical calculations beyond the critical state.

The general expressions of eigenfunctions $\boldsymbol{u}$ can be formulated as:

$$\begin{aligned}
u_x &= \tilde{u}_x(z)\left[A_1 \sin(mx)\sin(ky) + A_2 \sin(mx)\cos(ky) + A_3 \cos(mx)\sin(ky) + A_4 \cos(mx)\cos(ky)\right] \\
u_y &= \tilde{u}_y(z)\left[B_1 \sin(mx)\sin(ky) + B_2 \sin(mx)\cos(ky) + B_3 \cos(mx)\sin(ky) + B_4 \cos(mx)\cos(ky)\right] \\
u_z &= \tilde{u}_z(z)\left[C_1 \sin(mx)\sin(ky) + C_2 \sin(mx)\cos(ky) + C_3 \cos(mx)\sin(ky) + C_4 \cos(mx)\cos(ky)\right]
\end{aligned} \quad (5a)$$

They are to be considered in conjunction with the incompressiblity constraints

$$\begin{aligned}
A_1 m \tilde{u}_x(z) - B_4 k \tilde{u}_y(z) + C_3 \frac{d\tilde{u}_z(z)}{dz} &= 0 & A_2 m \tilde{u}_x(z) + B_3 k \tilde{u}_y(z) + C_4 \frac{d\tilde{u}_z(z)}{dz} &= 0 \\
-A_3 m \tilde{u}_x(z) - B_2 k \tilde{u}_y(z) + C_1 \frac{d\tilde{u}_z(z)}{dz} &= 0 & -A_4 m \tilde{u}_x(z) + B_1 k \tilde{u}_y(z) + C_2 \frac{d\tilde{u}_z(z)}{dz} &= 0
\end{aligned} \quad (5b)$$

as well as the Navier slip boundary conditions. In component form, the boundary conditions for $\tilde{\boldsymbol{u}}$ are

$$\tilde{u}_z(\pm 1) = 0 \quad .$$

$$\begin{aligned}
l_S \left.\frac{\partial \tilde{u}_x(z)}{\partial z}\right|_{z=1} + \tilde{u}_x(1) &= 0, & l_S \left.\frac{\partial \tilde{u}_x(z)}{\partial z}\right|_{z=-1} - \tilde{u}_x(-1) &= 0 \\
l_S \left.\frac{\partial \tilde{u}_y(z)}{\partial z}\right|_{z=1} + \tilde{u}_y(1) &= 0, & l_S \left.\frac{\partial \tilde{u}_y(z)}{\partial z}\right|_{z=-1} - \tilde{u}_y(-1) &= 0
\end{aligned} \quad (5c)$$

The $A_{1,2,3,4}$, $B_{1,2,3,4}$, $C_{1,2,3,4}$ are constants for delineating the periodic solution patterns in the $xy$ plane. The derivation of the 3D HMs is given in SM Part I [28]. Below we present the 3D HMs by dividing them into two separate symmetry categories. As already noted, the symmetry of the solution denotes the parity of $u_x, u_y$ with respect to the $z = 0$ plane. However, $u_z$ must have the opposite parity to that of $u_x, u_y$ as dictated by the incompressibility condition. This can be easily seen by expanding the *z dependence* of $\frac{\partial u_x}{\partial x}, \frac{\partial u_y}{\partial y}$ and $\frac{\partial u_z}{\partial z}$ around $z = 0$ and matching orders in $\Delta z$. Only if $u_z$ has the opposite parity to that of $u_x, u_y$, can there be matching orders for the possibility of satisfying the incompressibility condition.

The *antisymmetric* HMs are as follows:

$$u_x = \left[\sin(\mu z) - \frac{l_s \mu \cos(\mu) + \sin(\mu)}{l_s \nu \cosh(\nu) + \sinh(\nu)} \sinh(\nu z)\right]$$
$$\cdot \left[A_1 \sin(mx)\sin(ky) + A_2 \sin(mx)\cos(ky) + A_3 \cos(mx)\sin(ky) + A_4 \cos(mx)\cos(ky)\right]$$

$$u_y = \frac{k}{m}\left[\sin(\mu z) - \frac{l_s \mu \cos(\mu) + \sin(\mu)}{l_s \nu \cosh(\nu) + \sinh(\nu)} \sinh(\nu z)\right] \quad (6a)$$
$$\cdot \left[-A_4 \sin(mx)\sin(ky) + A_3 \sin(mx)\cos(ky) + A_2 \cos(mx)\sin(ky) - A_1 \cos(mx)\cos(ky)\right]$$

$$u_z = \frac{\nu}{m} \frac{l_s \mu \cos(\mu) + \sin(\mu)}{l_s \nu \cosh(\nu) + \sinh(\nu)} \left[\frac{\cosh(\nu)}{\cos(\mu)}\cos(\mu z) - \cosh(\nu z)\right]$$
$$\cdot \left[A_3 \sin(mx)\sin(ky) + A_4 \sin(mx)\cos(ky) - A_1 \cos(mx)\sin(ky) - A_2 \cos(mx)\cos(ky)\right]$$

As HMs must display the translational symmetry in the *xy* plane, $A_1, A_2, A_3, A_4$ are the constants for expressing the four independent degrees of freedom for adjusting solution's phase in the lateral plane.

For the $A_1$ branch,

$$p = \frac{\lambda}{m} \frac{l_s \mu \cos(\mu) + \sin(\mu)}{l_s \nu \cosh(\nu) + \sinh(\nu)} \sinh(\nu z) \cos(mx)\sin(ky) . \quad (6b)$$

Here $m, k, \mu$ are the wavevectors characterizing the eigenfunctions along the $x, y, z$ axis, respectively. For each given pair of wavevector $(m, k)$, we can obtain a countably infinite number of $\{\mu_n, n \in N^+\}$ from the following corresponding dispersion relation:

$$\nu[l_s \nu + \tanh(\nu)] + \mu[l_s \mu + \tan(\mu)] = 0, \quad \nu = \sqrt{m^2 + k^2}, \quad \mu = \sqrt{\lambda \operatorname{Re} - m^2 - k^2}. \quad (6c)$$

The *symmetric* HMs can be expressed similarly as

$$u_x = \left[\cos(\mu z) + \frac{l_s \mu \sin(\mu) - \cos(\mu)}{l_s \nu \sinh(\nu) + \cosh(\nu)} \cosh(\nu z)\right]$$
$$\cdot \left[A_1 \sin(mx)\sin(ky) + A_2 \sin(mx)\cos(ky) + A_3 \cos(mx)\sin(ky) + A_4 \cos(mx)\cos(ky)\right]$$

$$u_y = \frac{k}{m}\left[\cos(\mu z) + \frac{l_s \mu \sin(\mu) - \cos(\mu)}{l_s \nu \sinh(\nu) + \cosh(\nu)} \cosh(\nu z)\right]$$
$$\cdot \left[-A_4 \sin(mx)\sin(ky) + A_3 \sin(mx)\cos(ky) + A_2 \cos(mx)\sin(ky) - A_1 \cos(mx)\cos(ky)\right]$$

$$u_z = \frac{\nu}{m} \frac{l_s \mu \sin(\mu) - \cos(\mu)}{l_s \nu \sinh(\nu) + \cosh(\nu)}\left[-\frac{\sinh(\nu)}{\sin(\mu)}\sin(\mu z) + \sinh(\nu z)\right]$$
$$\cdot \left[A_3 \sin(mx)\sin(ky) + A_4 \sin(mx)\cos(ky) - A_1 \cos(mx)\sin(ky) - A_2 \cos(mx)\cos(ky)\right]$$

(7a)

For the $A_1$ branch,

$$p = -\frac{\lambda}{m}\frac{l_s \mu \sin(\mu) - \cos(\mu)}{l_s \nu \sinh(\nu) + \cosh(\nu)} \cosh(\nu z) \cos(mx)\sin(ky). \tag{7b}$$

The corresponding dispersion relation is given by

$$\nu\left[l_s \nu + \frac{1}{\tanh(\nu)}\right] + \mu\left[l_s \mu - \frac{1}{\tan(\mu)}\right] = 0, \ \nu = \sqrt{m^2 + k^2}, \ \mu = \sqrt{\lambda\,\text{Re} - m^2 - k^2}. \tag{7c}$$

In Eqs. (6,7), $\nu$ represents the wavevector in the *xy* plane, whereas $\mu$ is the wavevector along the *z* direction, which is also the eigenvector of the HMs, to be solved from the dispersion relation. It is to be noted that the eigenvalue $\lambda$, i.e., the decay time of the HM, is always bundled together with Re. Therefore the change in the value of Re can be completely compensated by the decay time of the HMs, without affecting the spatial configurations of the HMs.

The complete HM expression has the form $\boldsymbol{u}(\boldsymbol{r},0)e^{-\lambda t}$, where $\lambda$ is a real positive number ensured by the self-adjoint property of the NS equation [29,30]. Eigenfunction's time variation is important because the force balance of HMs is between the viscous force and the inertial acceleration, as well as the pressure field associated with each HM's velocity pattern. In contrast, for the Poiseuille flow the force balance is between the viscous force and the externally applied pressure gradient. All the 3D HMs satisfy both the incompressibility condition and the boundary conditions, as well as the linearized NS equation. These are explicitly verified in SM Part I [28]. As the basis set for expanding the channel flow velocity, the HMs are consistent with the channel geometry and boundary condition, in contrast to the Fourier basis [31]. The latter is more suitable for the infinite system with no boundaries.

For completeness, below we also write down the 1D and 2D HMs. For details, see SM Parts I and II [28]. The solution method for $\mu$ is detailed in SM Part III [28].

The 1D *antisymmetric* HMs have the form:

$$u_x(\mathbf{r}) = \sin(\mu z), \quad u_y(\mathbf{r}) = u_z(\mathbf{r}) = 0 \quad \text{or}$$
$$u_y(\mathbf{r}) = \sin(\mu z), \quad u_x(\mathbf{r}) = u_z(\mathbf{r}) = 0 \tag{8a}$$

Dispersion relation: $\quad l_s \mu + \tan(\mu) = 0, \quad p = \text{constant}, \quad \lambda \operatorname{Re} = \mu^2$

The 1D *symmetric* HMs have the form:

$$u_x(\mathbf{r}) = \cos(\mu z), \quad u_y(\mathbf{r}) = u_z(\mathbf{r}) = 0 \quad \text{or}$$
$$u_y(\mathbf{r}) = \cos(\mu z), \quad u_x(\mathbf{r}) = u_z(\mathbf{r}) = 0 \tag{8b}$$

Dispersion relation: $\quad \cot(\mu) - l_s \mu = 0, \quad p = \text{constant}, \quad \lambda \operatorname{Re} = \mu^2$

The 2D *antisymmetric* HMs have the form:

$$u_x = \left[ \sin(\mu z) - \frac{l_s \mu \cos(\mu) + \sin(\mu)}{l_s m \cosh(m) + \sinh(m)} \sinh(mz) \right] \left[ A_1 \sin(mx) + A_2 \cos(mx) \right]$$

$$u_z = \frac{l_s \mu \cos(\mu) + \sin(\mu)}{l_s m \cosh(m) + \sinh(m)} \left[ \frac{\cosh(m)}{\cos(\mu)} \cos(\mu z) - \cosh(mz) \right] \left[ A_2 \sin(mx) - A_1 \cos(mx) \right]$$

$$p = -\frac{\lambda}{m} \frac{l_s \mu \sin(\mu) - \cos(\mu)}{l_s m \sinh(m) + \cosh(m)} \sinh(mz) \left[ A_2 \sin(mx) - A_1 \cos(mx) \right]$$

$$\lambda \operatorname{Re} l_s + m \tanh(m) + \mu \tan(\mu) = 0, \quad m^2 + \mu^2 = \lambda \operatorname{Re}$$

or

$$u_y = \left[ \sin(\mu z) - \frac{l_s \mu \cos(\mu) + \sin(\mu)}{l_s k \cosh(k) + \sinh(k)} \sinh(kz) \right] \left[ A_1 \sin(ky) + A_2 \cos(ky) \right]$$

$$u_z = \frac{l_s \mu \cos(\mu) + \sin(\mu)}{l_s k \cosh(k) + \sinh(k)} \left[ \frac{\cosh(k)}{\cos(\mu)} \cos(\mu z) - \cosh(kz) \right] \left[ A_2 \sin(ky) - A_1 \cos(ky) \right]$$

$$p = -\frac{\lambda}{k} \frac{l_s \mu \sin(\mu) - \cos(\mu)}{l_s k \sinh(k) + \cosh(k)} \sinh(kz) \left[ A_2 \sin(ky) - A_1 \cos(ky) \right] \tag{9a}$$

$$\lambda \operatorname{Re} l_s + k \tanh(k) + \mu \tan(\mu) = 0, \quad k^2 + \mu^2 = \lambda \operatorname{Re}$$

The 2D *symmetric* HMs have the form

$$u_x = \left[\cos(\mu z) + \frac{l_S \mu \sin(\mu) - \cos(\mu)}{l_S m \sinh(m) + \cosh(m)} \cosh(mz)\right]\left[A_1 \sin(mx) + A_2 \cos(mx)\right]$$

$$u_z = \frac{l_S \mu \sin(\mu) - \cos(\mu)}{l_S m \sinh(m) + \cosh(m)}\left[-\frac{\sinh(m)}{\sin(\mu)}\sin(\mu z) + \sinh(mz)\right]\left[A_2 \sin(mx) - A_1 \cos(mx)\right],$$

$$p = -\frac{\lambda}{m}\frac{l_S \mu \sin(\mu) - \cos(\mu)}{l_S m \sinh(m) + \cosh(m)} \cosh(mz)\left[A_1 \cos(mx) - A_2 \sin(mx)\right]$$

$$\lambda \operatorname{Re} l_S + m\coth(m) - \mu\cot(\mu) = 0, \quad m^2 + \mu^2 = \lambda \operatorname{Re}$$

or

$$u_y = \left[\cos(\mu z) + \frac{l_S \mu \sin(\mu) - \cos(\mu)}{l_S k \sinh(k) + \cosh(k)} \cosh(kz)\right]\left[A_1 \sin(ky) + A_2 \cos(ky)\right]$$

$$u_z = \frac{l_S \mu \sin(\mu) - \cos(\mu)}{l_S k \sinh(k) + \cosh(k)}\left[-\frac{\sinh(k)}{\sin(\mu)}\sin(\mu z) + \sinh(kz)\right]\left[A_2 \sin(ky) - A_1 \cos(ky)\right] \quad . \tag{9b}$$

$$p = -\frac{\lambda}{k}\frac{l_S \mu \sin(\mu) - \cos(\mu)}{l_S k \sinh(k) + \cosh(k)} \cosh(kz)\left[A_1 \cos(ky) - A_2 \sin(ky)\right]$$

$$\lambda \operatorname{Re} l_S + k\coth(k) - \mu\cot(\mu) = 0, \quad k^2 + \mu^2 = \lambda \operatorname{Re}$$

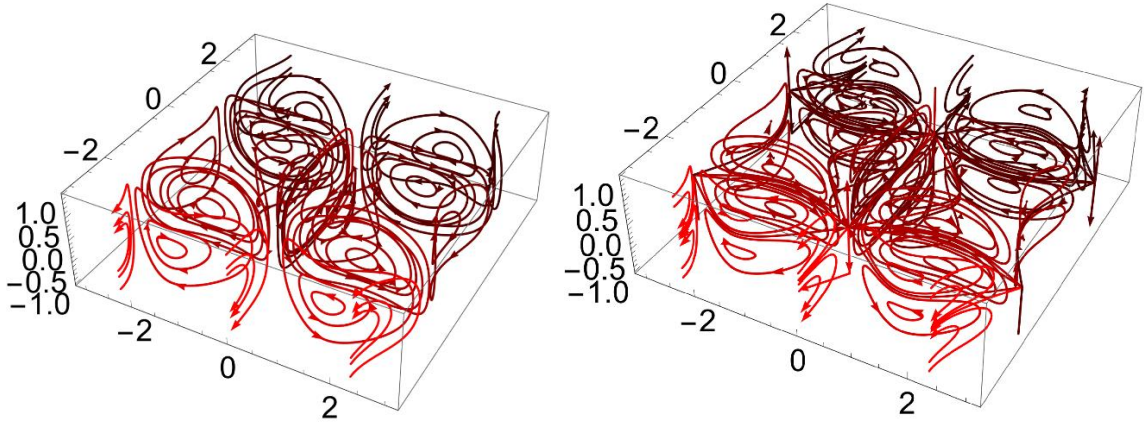

(a)  λ=0.000934 m=1.009` k=1.009`   (b)  λ=0.0021 m=1.009` k=1.009`

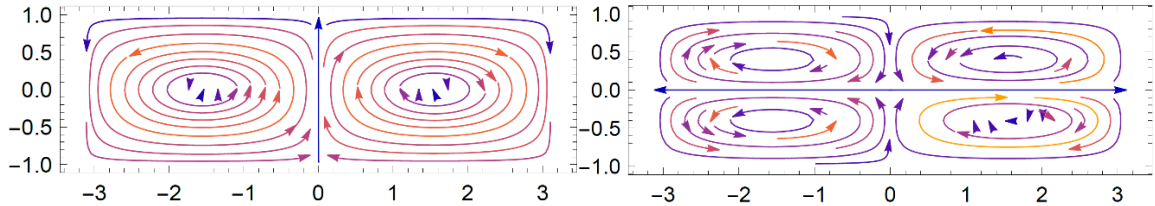

(c)  λ=0.000931 m=1.009`   (d)  λ=0.00206 m=1.009`

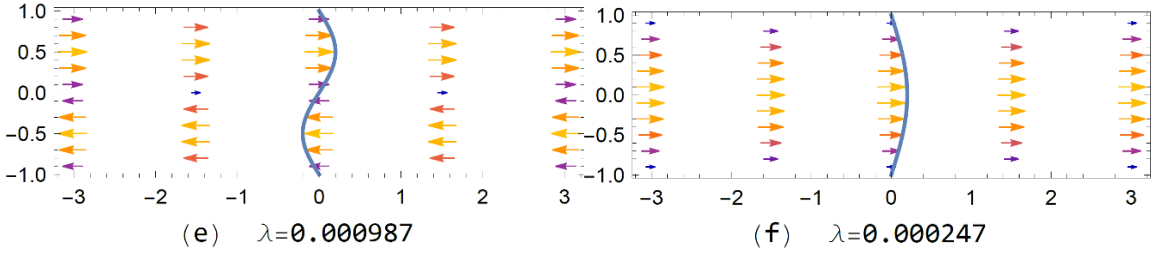

**FIG. 1**: A schematic illustration of **(a)** antisymmetric 3D HM, **(b)** symmetric 3D HM, **(c)** antisymmetric 2D HM, **(d)** symmetric 2D HM, **(e)** antisymmetric 1D HM, and **(f)** symmetric 1D HM. The relevant eigenvalue and eigenvector values are below each figure.

In Fig. 1, we graphically illustrate the 3D, 2D, and 1D HMs with Re=$10^4$. Among all the 1D, 2D and 3D HMs, the 1D symmetric HMs are special because they display *center of mass motion*, which is crucial for giving rise to the change in the net channel flow rate beyond the critical state (Re$_c$, $m_c$). In contrast to the isotropic infinite systems in which the center of mass motion cannot be allowed due to the absence of external force, there can be two external forces in the channel flow: the frictional reaction force from the boundaries, and the force from the externally applied pressure gradient. According to Newton's law, the external forces can alter the center of mass momentum in the channel flow. Beyond the critical state, nonlinear couplings between the HMs with zero net momentum, and the 1D symmetric modes with net flow, are shown to generate boundary reaction force in a direction opposite to the pressure force, leading to a decrease in the net flow. An account of the energy conservation and force balance under this scenario is given in the last section.

The HMs completely separate the NS equations into the spatial and temporal components. They can describe all possible fluid particles' motions in the spatial domain [17,18,24] from the continuum perspective. Each HM satisfies both the linearized NS equation and the incompressibility condition, and the relevant pressure field is coupled to the velocity field through the pressure Laplace (Poisson) equation [32], with the boundary condition given by $\left[ \partial p / \partial z = (1/\text{Re}) \partial^2 u_z / \partial z^2 \right]_{z=\pm 1}$.

It is shown below that the pressure gradient term can be eliminated when the NS equation is reduced to a coupled set of time-varying autonomous ODEs, by expanding the velocity in terms of the HMs. Hence the pressure gradient term is not involved either in time-evolving the autonomous ODEs (SM Part IV [28]), or in analyzing the critical state (Re$_c$, $m_c$).

In order to simplify the classification scheme, below we introduce a unified expression for the HMs that includes all the cases stated above.

$$u_x = \left(m+\delta_{d,0}\right) \left\{ \begin{array}{l} \delta_{s,0}\left[\left(\delta_{\kappa,0}\delta_{d,0}+\delta_{d,1}\right)\sin(\mu z)-\delta_{d,1}\dfrac{l_s\mu\cos(\mu)+\sin(\mu)}{l_s\nu\cosh(\nu)+\sinh(\nu)}\sinh(\nu z)\right] \\ +\delta_{s,1}\left[\left(\delta_{\kappa,0}\delta_{d,0}+\delta_{d,1}\right)\cos(\mu z)+\delta_{d,1}\dfrac{l_s\mu\sin(\mu)-\cos(\mu)}{l_s\nu\sinh(\nu)+\cosh(\nu)}\cosh(\nu z)\right] \end{array} \right\}$$
$$\cdot \left[\delta_{d,1}\left(\delta_{o_x,0}\sin(mx)+\delta_{o_x,1}\cos(mx)\right)\left(\delta_{o_y,0}\sin(ky)+\delta_{o_y,1}\cos(ky)\right)+\delta_{d,0}\right]$$

$$u_y = \left(k+\delta_{d,0}\right) \left\{ \begin{array}{l} \delta_{s,0}\left[\left(\delta_{\kappa,1}\delta_{d,0}+\delta_{d,1}\right)\sin(\mu z)-\delta_{d,1}\dfrac{l_s\mu\cos(\mu)+\sin(\mu)}{l_s\nu\cosh(\nu)+\sinh(\nu)}\sinh(\nu z)\right] \\ +\delta_{s,1}\left[\left(\delta_{\kappa,1}\delta_{d,0}+\delta_{d,1}\right)\cos(\mu z)+\delta_{d,1}\dfrac{l_s\mu\sin(\mu)-\cos(\mu)}{l_s\nu\sinh(\nu)+\cosh(\nu)}\cosh(\nu z)\right] \end{array} \right\}$$
$$\cdot \left[\delta_{d,1}\left(\delta_{o_x,1}\sin(mx)-\delta_{o_x,0}\cos(mx)\right)\left(-\delta_{o_y,1}\sin(ky)+\delta_{o_y,0}\cos(ky)\right)+\delta_{d,0}\right]$$

$$u_z = \nu \left\{ \begin{array}{l} \delta_{s,0}\dfrac{l_s\mu\cos(\mu)+\sin(\mu)}{l_s\nu\cosh(\nu)+\sinh(\nu)}\left[\dfrac{\cosh(\gamma)}{\cos(\mu)}\cos(\mu z)-\cosh(\nu z)\right] \\ +\delta_{s,1}\dfrac{l_s\mu\sin(\mu)-\cos(\mu)}{l_s\nu\sinh(\nu)+\cosh(\nu)}\left[-\dfrac{\sinh(\nu)}{\sin(\mu)}\sin(\mu z)+\sinh(\nu z)\right] \end{array} \right\}$$
$$\cdot \delta_{d,1}\left(\delta_{o_x,1}\sin(mx)-\delta_{o_x,0}\cos(mx)\right)\left(\delta_{o_y,0}\sin(ky)+\delta_{o_y,1}\cos(ky)\right)$$

(10)

Here $\delta$ is the Kronecker delta function and $s, \kappa, d, o_x, o_y$ are the 5 hyper-indices used to denote different branches of the HMs. Values of $s, \kappa, d, o_x, o_y \in \{0 \text{ or } 1\}$. Here $s = 0$ denotes the antisymmetric HMs and $s = 1$ denotes the symmetric HMs; $\kappa = 0$ denotes the $x$ branch and $\kappa = 1$ denotes the $y$ branch; $d = 0$ denotes the 1D HMs and $d = 1$ denotes the 2D or 3D HMs depending on the values of $(m, k)$ tuple; $o_x = 0$ denotes the $sin(mx)$ phase term of $u_x$ and $o_x = 1$ denoting the $cos(mx)$ phase term of $u_x$; and $o_y = 0$ denotes the $sin(ky)$ phase term of $u_y$, and $o_y = 1$ denotes the $cos(ky)$ phase term of $u_y$.

As shown in Fig. 1, except for the 1D symmetric HMs, the remaining HMs have zero net momentum and zero net angular momentum within each periodic domain, mostly comprising clockwise and counterclockwise pairs of vortices. When used in calculations, every HM, denoted below as $u_\alpha$, is normalized so that its norm is 1. Normalization is over the volume specified by $\left(x = \pm\dfrac{1}{2}\sqrt{A}, y = \pm\dfrac{1}{2}\sqrt{A}, z = \pm 1\right)$, where $A$ denotes the area in the $xy$ plane covered by the volume integration.

## III. Velocity expansion and the reduction of the Navier-Stokes equation to an autonomous system of ODEs

For the channel flow, we choose the unperturbed steady state, i.e., the Poiseuille flow $\boldsymbol{u}^P(\boldsymbol{r})$, as the zeroth-order ground state. The full nonlinear perturbed NS equation takes the form

$$\frac{\partial \boldsymbol{u}}{\partial t} = -\left(\boldsymbol{u}^P(\mathbf{r})\cdot\nabla\right)\boldsymbol{u} - (\boldsymbol{u}\cdot\nabla)\boldsymbol{u}^P(\mathbf{r}) - \nabla p + \frac{1}{\text{Re}}\nabla^2\boldsymbol{u} - (\boldsymbol{u}\cdot\nabla)\boldsymbol{u}, \tag{11}$$

where the total velocity field is reminded to be $\boldsymbol{v}(\boldsymbol{r},t) = \boldsymbol{u}(\boldsymbol{r},t) + \boldsymbol{u}^P(\boldsymbol{r})$. By expanding $\boldsymbol{u}(\boldsymbol{r},t)$ in Eq. (11) with the HMs of Eq. (10), we reduce Eq. (11) to a system of coupled autonomous equations for the vector of expansion coefficients $\boldsymbol{c}(t)$ in which $t$ is the only independent variable. Here we have

$$\begin{aligned}
\boldsymbol{u}(\mathbf{r},t) &= \sum_{\alpha=1}^{\infty} c_\alpha(t)\boldsymbol{u}_\alpha(\mathbf{r}), \quad c_\alpha(t)\in\mathbf{R},\ \boldsymbol{u}_\alpha(\mathbf{r})\in\mathbf{R}^3 \\
\frac{\partial c_\gamma(t)}{\partial t} &= -\sum_{\alpha=1}^{\infty}\left\langle \boldsymbol{u}_\gamma(\mathbf{r}), \left(\boldsymbol{u}^P(\mathbf{r})\cdot\nabla\right)\boldsymbol{u}_\alpha(\mathbf{r})\right\rangle c_\alpha(t) - \sum_{\alpha=1}^{\infty}\left\langle \boldsymbol{u}_\gamma(\mathbf{r}),(\boldsymbol{u}_\alpha(\mathbf{r})\cdot\nabla)\boldsymbol{u}^P(\mathbf{r})\right\rangle c_\alpha(t) \\
&\quad -\left\langle \boldsymbol{u}_\gamma(\mathbf{r}),\nabla p\right\rangle + \frac{1}{\text{Re}}\sum_{\alpha=1}^{\infty}\left\langle \boldsymbol{u}_\gamma(\mathbf{r}),\nabla^2\boldsymbol{u}_\alpha(\mathbf{r})\right\rangle c_\alpha(t) \\
&\quad -\sum_{\alpha=1}^{\infty}\sum_{\beta=1}^{\infty}\left\langle \boldsymbol{u}_\gamma(\mathbf{r}),(\boldsymbol{u}_\alpha(\mathbf{r})\cdot\nabla)\boldsymbol{u}_\beta(\mathbf{r})\right\rangle c_\alpha(t)c_\beta(t)
\end{aligned}. \tag{12}$$

Here $\{\boldsymbol{u}_\alpha(\mathbf{r})\}$ denotes the set of complete, normalized HMs with subscript $\alpha$ as the labeling index, $\langle\ \rangle$ denotes inner product, and all the spatially-dependent variables are converted into analytical scalar expressions through inner product projections. Equation (12) is a first order system of coupled ODEs. They can be written in the matrix form as

$$\frac{\partial \boldsymbol{c}(t)}{\partial t} = \mathcal{L}\cdot\boldsymbol{c}(t) - \vec{\mathcal{N}}[\boldsymbol{c}(t)]. \tag{13a}$$

Here $\mathcal{L}$ represents the coefficients matrix of the linear terms and $\vec{\mathcal{N}}$ represents the 3rd order tensor term. By utilizing the incompressibility property and periodicity of the HMs in the *xy* plane, *the pressure gradient term can be eliminated* in Eq. (12). For details, see SM Part IV [28]. The NS equation is thereby transformed formally into a system of coupled ODEs:

$$\frac{\partial c_\gamma}{\partial t} = \sum_{\alpha=1}^{\infty} \mathcal{L}_{\gamma\alpha} c_\alpha - \sum_{\alpha=1}^{\infty}\sum_{\beta=1}^{\infty} \vec{\mathcal{N}}_{\alpha\beta\gamma} c_\alpha c_\beta$$

$$\mathcal{L}_{\gamma\alpha} = -\langle \boldsymbol{u}_\gamma(\mathbf{r}), (\boldsymbol{u}^P(\mathbf{r})\cdot\nabla)\boldsymbol{u}_\alpha(\mathbf{r})\rangle - \langle \boldsymbol{u}_\gamma(\mathbf{r}), (\boldsymbol{u}_\alpha(\mathbf{r})\cdot\nabla)\boldsymbol{u}^P(\mathbf{r})\rangle + \frac{1}{\text{Re}}\langle \boldsymbol{u}_\gamma(\mathbf{r}), \nabla^2 \boldsymbol{u}_\alpha(\mathbf{r})\rangle \quad . \quad (13b)$$

$$\vec{\mathcal{N}}_{\alpha\beta\gamma} = \langle \boldsymbol{u}_\gamma(\mathbf{r}), (\boldsymbol{u}_\alpha(\mathbf{r})\cdot\nabla)\boldsymbol{u}_\beta(\mathbf{r})\rangle$$

The two operators $\mathcal{L}, \vec{\mathcal{N}}$ are only functions of Re and slip length $l_s$. It should be noted that the Poiseuille flow enters only in the matrix elements of the linear term, $\mathcal{L}$, and is absent in $\vec{\mathcal{N}}$. Theoretically we should be able to obtain all the channel fluid dynamics information by time-evolving the above set of coupled ODEs. Here we choose the explicit Runge-Kutta scheme of 4$^{th}$ order accuracy [33,34] to numerically compute the time evolution trajectories of Eq. (13). Details of the numerical scheme are presented in SM Part V [28]. By choosing an appropriate time step $\Delta t$, the numerical errors can be controlled.

For the linear operator $\mathcal{L}$, the pressure corresponding to each $\boldsymbol{u}_\alpha$ can be recovered by solving a pressure-Poisson equation

$$\nabla^2 p_\alpha = -\nabla \cdot \left[ \left(\boldsymbol{u}^P(\mathbf{r})\cdot\nabla\right)\boldsymbol{u}_\alpha + (\boldsymbol{u}_\alpha \cdot \nabla)\boldsymbol{u}^P(\mathbf{r}) \right] = 4z\frac{\partial (\boldsymbol{u}_\alpha)_x}{\partial x} \quad , \quad (13c)$$

with the boundary condition

$$\left.\frac{\partial p_\alpha}{\partial z}\right|_{z=\pm 1} = \frac{1}{\text{Re}}\left.\frac{\partial^2 u_{\alpha,z}}{\partial z^2}\right|_{z=\pm 1}. \quad (13d)$$

In anticipation of later developments, we note that the exponential decay of each HM as a function of time is implicitly incorporated in Eq. (13) through the matrix/tensor elements that contain the (viscous dissipation of its) spatial velocity pattern of the HMs. The sustained presence of the HMs in the flow pattern, as will be shown to be the case beyond the critical state, can only arise when there is sustaining work done by the externally supplied pressure. A detailed consideration of the energy conservation law is given below.

## IV. Evaluation of the critical state parameters

The stability of the Poiseuille flow under the nonslip boundary condition is characterized by the existence of a critical value Re$_c$ [8-10], so that when Re<Re$_c$ the Poiseuille flow is stable under small perturbations; while the perturbed velocity field can increase exponentially for Re>Re$_c$, accompanied by the emergence of a critical wavevector. By considering infinitesimal perturbed velocity field $\boldsymbol{u}(\boldsymbol{r},t)$ in the linear regime, Eq. (13b) may be expressed as

$$\frac{\partial c_\gamma}{\partial t} = \sum_{\alpha=1}^{\infty} \mathcal{L}_{\gamma\alpha} c_\alpha. \tag{14}$$

Equation (14) represents a linear system whose dynamic stability is determined by the maximum real part of the linear operator spectrum. The critical (Re$_c$, $m_c$) corresponds to the value of (Re, $m$) when one of $\mathcal{L}$'s eigenvalues' real part first crosses 0. The perturbations decay exponentially below the critical state and increase exponentially otherwise. In the search for the critical state parameter values, we have fixed $k$=0, since from the literature [35] it is already known that the most unstable mode can be found in the 2D case with $k$=0. This point is further elaborated below. We have numerically calculated the linear operator's spectrum from its matrix formulation $\mathcal{L}_{\alpha\gamma}$ as given in Eq. (13b). Here we note that the $\mathcal{L}_{\alpha\gamma}$ matrix has a block-diagonal structure, in which each $2N \times 2N$ block corresponds to a unique value of $m$ ($k$=0), with $N$ being the number of $\mu$'s used in the search. By fixing the number of $\mu$'s at 400, with $\Delta m = 0.0001$ as the step size in the $m$ scan, we focused on the subspace spanned by the finite number of HMs:

$$\frac{\partial c_\gamma}{\partial t} \approx \sum_{\alpha=1}^{800} \mathcal{L}_{\gamma\alpha} c_\alpha. \tag{15}$$

Bisection method was employed to locate the precise value of Re$_c$. By denoting the maximum real part of the eigenvalues of $\mathcal{L}_{\gamma\alpha}$ as Real($\sigma$) and the imaginary part as Imag($\sigma$), we scanned the (Re, $m$) values to determine the first sign change of Real($\sigma$), and denote the corresponding (Re$_c$, $m_c$) as the critical state. We compare our result with that of Orszag [8]—Re$_c$ = 5772.22, $m_c$ = 1.02056. Our formalism yields Re$_c$ = 5772.2 and $m_c$ = 1.0203. Excellent agreement is seen. It should be noted that Imag($\sigma_c$) = 0.269 also agrees well with the Orszag result, and represents an oscillation frequency of the critical state. A video illustration of this oscillating critical state is shown in SM Part IV [28].

For the 2-dimensional channel flow, Orr-Sommerfeld stability equation [8,10,35] utilizes the nonslip boundary conditions to fix the normal derivative $\partial u_z/\partial n$=0. In this way, the perturbed NS equation can be compressed into a 1D, 4$^{th}$ ODE. The aim is to locate the value of Re when Real($\sigma$) = 0. Furthermore, if the eigen wavevector $k \neq 0$, then under the nonslip boundary condition the Squire transformation [35] shows that any 3D channel flow's perturbation is equivalent to a specific 2D channel flow perturbation. Hence the task of obtaining the 3D channel flow's critical state is transformed into solving for the (Re$_c$, $m_c$) of its corresponding 2D flow. By specifying a given input $k \neq 0$, the critical $m_c$ as well as its corresponding Re$_c$ can be uniquely determined through the Squire transformation. In Fig. 2 we plot the Re$_c$ as a function of $m_c, k$. The conventional definition of $Re_c$ corresponds to the minimum value of the curve at $k=0$, $m_c = 1.02$, which is the most unstable point and labeled as a red dot in Fig. 2. As anticipated previously, $m_c$=1.02 defines a natural (inverse) length scale, which will be used in our discretization of the wavevectors in the *xy* plane for numerical evaluation of the channel flow beyond the critical state.

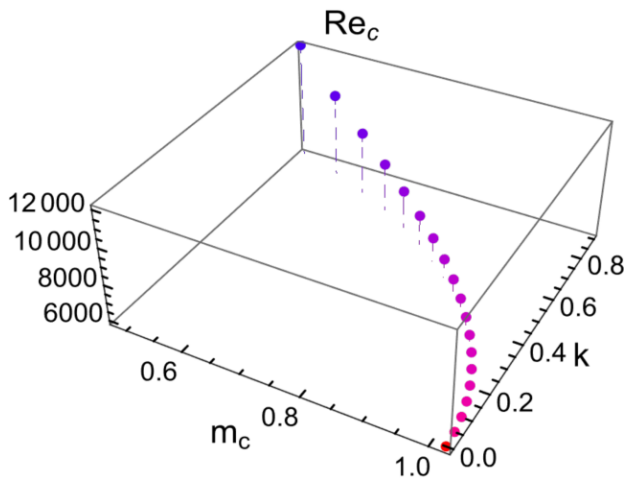

**FIG. 2**. Plot of $Re_c$ as a function of the eigen-wavevectors $m_c, k$. The conventional definition of $Re_c$=5772.2 is noted to be the minimum point of the discretized curve (i.e., the least stable mode), labeled by a red dot. Here *k* is an input parameter.

The Orr-Sommerfeld stability equation generally does not apply to the case when $l_s \neq 0$, but our formalism can easily deal with the more generalized case. No extra modifications are needed. In Fig. 3, the evaluated $(Re_c, m_c)$ are plotted as a function of slip length $l_s$ ranging from 0 to $14 \times 10^{-4}$. We see that there exists a minimum $l_s = 0.006$ at which the flow has the smallest $Re_c$. It describes an interesting behavior: Increase in the slip length will first lead the system to be less stable, and only after the minimal point is passed will the system become more stable with increasing slip. However, this behavior is conditioned on allowing the Poiseuille flow rate to increase with $l_s$ as given by Eq. (3b). If we switch to the condition of constant flow rate independent of $l_s$, then the variation of $Re_c$ with a finite slip length would revert to a monotonic behavior, in agreement with that shown in [21].

An aspect not previously observed is the actual configuration and dynamics of the critical state. Since in our case the critical state configuration is a by-product of the critical eigenvalue of the linear operator $\mathcal{L}$, its configuration can be easily constructed from the set $\{u_\alpha\}_c$. It turns out that while $\text{Real}(\sigma_c) = 0$, its imaginary part, $\text{Imag}(\sigma_c) = 0.269$ (for the nonslip case), represents an oscillation frequency for the critical state. In SM Part IV [28], we give a video illustration of the oscillating critical state that comprises a vortex and anti-vortex pair that oscillates at a time periodicity of $2\pi/\text{Imag}(\sigma_c) = 23.3$. Obviously, this oscillation behavior arises from the participation of the Poiseuille flow as the energy source, whose velocity profile has explicitly participated in the matrix elements of the linear operator $\mathcal{L}$ as shown in Eq. (13b).

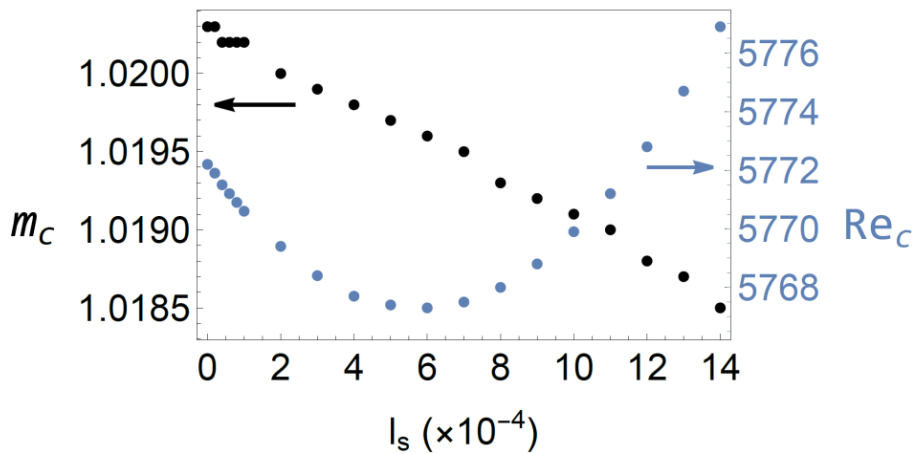

**FIG. 3.** A plot of the critical wavevector $m_c$ (in black) and the critical Reynolds number $\text{Re}_c$ (in blue), both as a function of the slip length $l_s$. Here the Poiseuille flow rate is allowed to vary with the slip length as stipulated by Eq. (3b).

## V. Channel flow beyond the critical state

### A. Initial state sampling

Thermal fluctuations constitute an inherent part of the physical system at equilibrium. They are manifest as a source of noise arising from the thermal bath. Without thermal excitations, the Poiseuille flow can be absolutely stable to any Reynolds number. In previous works [17,18], it was shown that in terms of the complete set of HMs, the fluctuation-dissipation theorem [23,24,36,37] may be expressed in a simple expression involving the average of the inverse of the HMs' eigenvalues. By considering the fluid molecular motions as superpositions of the HMs, the thermal fluctuations in fluid can be represented from the continuum perspective as canonical ensembles of the projection coefficients $c(t)$ with initial Gaussian distributions following the Heisenberg picture of phase space averages. Moreover, since each HM represents one degree of freedom [24,17,18], from the equipartition theorem each HM should have the same thermal energy [24]. Hence the initial probability density function of the projection coefficients $c(t)$ should obey the Gaussian distribution

$$f(\bm{c}) = \frac{1}{\varepsilon^N \sqrt{(2\pi)^N}} e^{-\frac{1}{2\varepsilon^2} \bm{c}^T \cdot \bm{c}}, \quad \bm{c} \in \mathbf{R}^N, \tag{16}$$

where $\varepsilon^2$ may be regarded as the thermal energy. Given the initial state with coefficients distributed in accordance to $f(\mathbf{c})$, the ensemble averaged perturbed net flow $\tilde{Q}(t)$ can be computed as

$$\tilde{Q}(t) = \int \mathbf{c}_{1D}^T(\mathbf{c}_0, t) \cdot \mathbf{q}_{1D} f(\mathbf{c}_0) d\mathbf{c}_0, \qquad \mathbf{q}_{1D} - 1D \text{ symmetric HMs flow rate}, \tag{17}$$

where $\mathbf{q}_{1D}$ is the vector of 1D symmetric HMs' flow rates.

To simulate the averaged flow rate under thermal excitations, we first prepare a set of independent and normally distributed random variables $\mathbf{c} \coloneqq \{c_1, \ldots, c_N\}$ with variance $\varepsilon^2$. Assuming that a total of $K$ trajectories are used to calculate the ensemble average, we repeat the above process $K$ times to obtain a set of random variable vector sets $C_0 = \{\mathbf{c}_1, \mathbf{c}_2, \ldots, \mathbf{c}_K\}$. Then by performing the time evolution stipulated by Eq. (13) with each element in $C_0$ set to be the initial state, we obtain $K$ independent trajectories $\{\mathbf{c}_1(t), \mathbf{c}_2(t), \ldots, \mathbf{c}_K(t)\}$. The ensemble-averaged flow rate at any time $t$ may be evaluated as

$$\tilde{Q}(t) \approx \frac{1}{K} \sum_{\gamma=1}^{K} \mathbf{c}_{\gamma,1D}^T(t) \cdot \mathbf{q}_{1D}. \tag{18}$$

In our numerical time evolution of Eq. (13b), we used $K = 10$ evolution trajectories, where the initial state of each trajectory comprises the 2D antisymmetric modes along the $x$ and $y$ branches with discretized values of $m$ and $k$ in multiples of $m_c \times 2^i$, where $i$=0,1, 2, 3, 4, 5. Multiples of 2 were chosen because they represent the maximum couplings mediated by the tensorial term [38]. The amplitudes of the modes were generated by the Gaussian random process described above, with $\varepsilon^2 = 0.04$. Only 2D antisymmetric modes were chosen as the initial states since the symmetric modes have not been found to be unstable, a fact that has not yet been rigorously proved. In our numerical evaluations, Re was fixed at $10^4$, and the computational discretization in the $x$, $y$ directions was set at $\Delta m = m_c$, so that the length $L$ and width $W$ of our computational domain in the $xy$ plane are fixed at $L=W = \pm\frac{\pi}{\Delta m} = \pm\frac{\pi}{m_c}$.

It should be noted that even though the initial states were all 2D antisymmetric modes, as time evolved, the 3D modes appeared. Hence the 3D HMs are absolutely needed in the numerical experiment. In the time evolution as stipulated by Eq. (13), a total of 6596 HMs were involved.

Poiseuille flow below the critical state is always stable with all excited HMs decaying exponentially as a function of $t$. Hence the ensemble-averaged *perturbed* net flow $\tilde{Q}(t)$ rapidly decays to 0 below the critical state. Above the critical point, the situation changes due to the amplification effect of the linear operator $\mathcal{L}$. However, the amplification effect cannot be sustained since the total kinetic energy is continuously being dissipated, while the other perturbation components are also excited through the nonlinear coupling term $(u \cdot \nabla)u$. In this manner, the exponential increase is suppressed.

What happens when an extremely long time $t$ passes? From the following simple argument we can see that the perturbed flow cannot be zero. Suppose at a time $t$ all the perturbation components tend to 0 above the critical state, then the linear analysis tells us that the thermal excitations will again make those HMs with positive real part of the eigenvalues exponentially increase in magnitude, and the whole process can start over again. It follows that the final equilibrium state should have a nonzero ensemble-averaged perturbed net flow, accompanied by fluctuations.

### B. Reduced net flow rate

Physically, this anticipated outcome originates from the excitation of the 1D symmetric modes mediated by the tensorial term, leading to counter flow in a direction opposite to the Poiseuille flow. This is indeed the result as shown in Fig. 4.

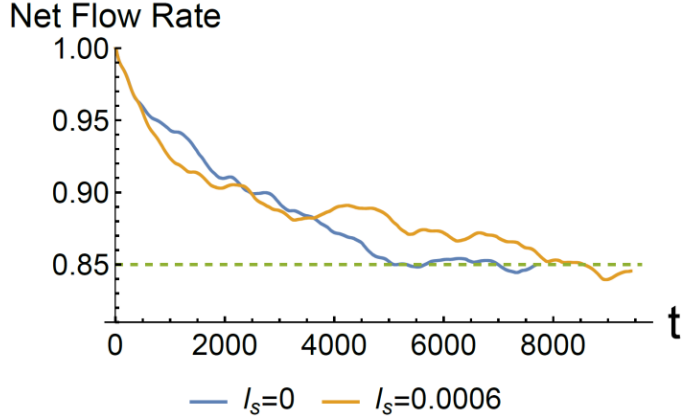

**FIG. 4.** A plot of the normalized net flow rate, in unit of $Q^p$, as a function of time $t$. The time evolution is evaluated from Eq. (13) and Eq. (18), with the Reynolds number set at Re=$10^4$. The ensemble-averaged net flow rate, with ten independently generated initial states, is seen to decrease as a function of time from the initial Poiseuille flow rate. The yellow curve, indicative of the case for the finite slip length, is seen to approach the equilibrium flow rate slower than that of the nonslip case.

We have adopted the 4$^{th}$ order explicit Runge-Kutta numerical scheme to evaluate the time evolution of Eq. (13). Details of the numerical scheme can be found in SM Part V [28]. In Fig. 4 we plot the time evolution of the flow rate variation as evaluated by Eq. (13) and (18), at Re=$10^4$. The curves were generated by ensemble-averaging ten independent initial states, run in parallel on 4 Amazon servers for about 3 months. The net flow rate shows the deviation from $Q^P$. We should be reminded that even though the Reynolds number used in the evaluation of Eq. (13) is above the critical value, the Poiseuille flow, *in the absence of perturbed modes*, should remain the same. This is because even though the pressure varies inversely as 1/Re, the decrease in viscosity exactly compensates the lowered pressure to maintain the same flow rate. Hence the deviation as shown in Fig. 4 can only be ascribed to the appearance of the perturbed HMs, especially the 1D symmetric modes. The net flow rate is seen to decrease until it reaches a plateau that is ~15% lower than the Poiseuille flow rate. However, not seen from Fig. 4 is the presence of the many HMs with vortices and anti-vortices that accompany this state. There are perceptible fluctuations around the plateau.

An interesting observation is that in accordance to Newton's law, the reduced net flow would require the boundary reaction force to be along the direction of the counter flow, i.e., against the Poiseuille flow. That would be opposite to what one would usually expect. This puzzle is resolved in the context of the counter flow profiles and force balance, presented below.

### C. Energy conservation law

With the solution of the velocity expansion coefficient set $\{c_\gamma\}$, we can also verify, at every time instant, the energy conservation for each independent trajectory as a validation of our numerical implementation. The energy conservation law states that

$$\frac{dE_{kinetic}(t)}{dt} = W_p(t) - W_d(t), \qquad (19a)$$

where $W_p(t)$ is the rate of energy input exerted by the external pressure, $W_d(t)$ is the rate of viscous dissipation, and the left hand side of Eq. (19a) is the rate of change of the kinetic energy. In SM Part VI [28] we give the derivation of the expressions for $E_{kinetic}(t), W_p(t)$ and $W_d(t)$ in terms of the HMs and the velocity expansion coefficients. This form of the energy conservation law can be expressed as

$$\frac{dE_{kinetic}}{dt} = \frac{1}{2}\frac{d}{dt}\sum_{\alpha=1}^{N} w_\alpha^2(t) = \frac{2}{\text{Re}} \times 4LW \times \left( \sum_{\substack{\alpha \in \{1\text{D symmetric} \\ \text{HMs index}\}}} w_\alpha(t) \int_{-1}^{1} (\boldsymbol{u}_\alpha)_x(\boldsymbol{r}) dz \right) - \sum_{\alpha=1}^{N} w_\alpha^2(t) \lambda_\alpha, \quad (19b)$$

where $w_\alpha(t)$ denotes the projection coefficients of the total velocity field $\boldsymbol{v}$, including the Poiseuille flow velocity. The latter is expanded in terms of the 1D symmetric HMs. Here *N*=6596 denotes the total number of HMs involved. Derivation of Eq. (19b) is given in SM

Part VI [28]. The first term on the right hand side represents the energy density input exerted by the external pressure, while the second term represents the energy dissipation rate caused by the viscous stress tensor. The energy conservation law tells us that the change in the total kinetic energy, $\Delta E_{kinetic}(t)$, must be equal to the net work. In Fig. 5 we plot the ensemble-averaged, time-integrated kinetic energy as a function of $t$, while the inset shows the same for one selected evolution trajectory. We see excellent consistency between the two curves as evaluated from two sides of Eq. (19b), indicating energy conservation to be well obeyed by our numerical scheme for every evolution trajectory.

It is clear from Fig. 5 that the general trend in the kinetic energy evolution is along a decreasing path. This trend is due to the fact that most of perturbation modes' energy becomes concentrated in the 1D symmetric modes as $t$ increases, with the rest of the HMs having at least an order of magnitude less energy. Since the effect of the 1D symmetric modes is to decrease the net flow, hence the kinetic energy of the whole system decreases. While the other zero-momentum modes (mostly in the form of coupled vortices and anti-vortices) have small amplitudes, they provide the energetic link between the 1D symmetric modes and the Poiseuille flow. Otherwise the 1D symmetric modes would decay and disappear.

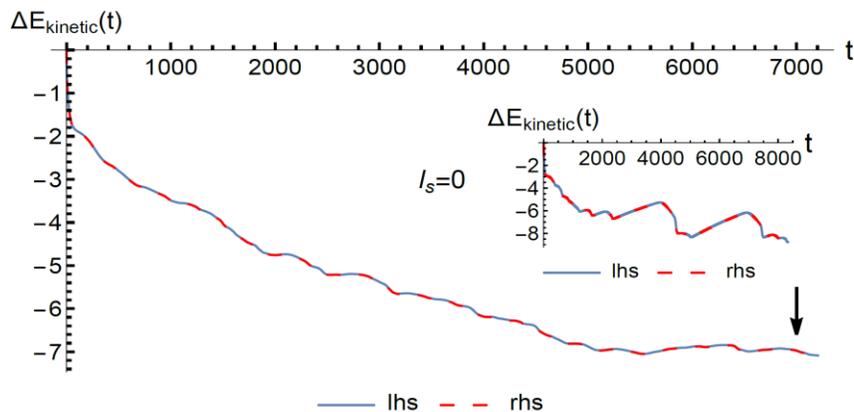

**FIG. 5.** The ensemble-averaged kinetic energy as evaluated from the energy conservation relation, Eq. (19), plotted as a function of time $t$. Here the kinetic energy is numerically time-integrated from Eq. (19). The blue line indicates the left hand side of Eq. (19), and the red dashed line indicates the right hand side. The nonslip boundary condition was used in this case. The same quantity for a single trajectory is plotted in the inset. It is seen that the energy conservation law is very well obeyed in the time evolution, and the kinetic energy decreases as a function of time, until it reaches a plateau. The vertical arrow indicates the time, $t=7000$, at which the flow profile is plotted in Fig. 6.

### D. Net flow profile and the counter flow

Beyond the critical Reynolds number the net flow comprises both the Poiseuille flow and the 1D symmetric modes. In Fig. 6(a) we plot the total velocity profile, $v_x$, at $t=0$ and $t=7000$ for a comparison. Here we consider only the nonslip boundary condition; the same holds true for what follows. The time instance $t=7000$ falls in what we denote as the "equilibrium state" time period. It is seen that the two velocity profiles are largely similar. However, the presence of the 1D symmetric modes can be made apparent by subtracting off the $t=0$ Poiseuille profile from the net flow profile. The result, denoted as the counter flow $v_x^C$, is plotted in Fig. 6(b) at three time instances. The $v_x^C$ is composed of different 1D symmetric modes, not all of them flowing along the same direction, e.g., at $t=7000$, $v_x^C = -0.158\cos(1.571z) + 0.01\cos(4.712z) + 0.0075\cos(7.854z) - 0.0074\cos(10.996z) + \cdots$. Hence one can see that in the vicinity of the solid boundary, there can be opposite slopes of $\left[\frac{\partial v_x^C}{\partial z}\right]_{z=\pm 1}$ that reflect the different directions of the boundary reaction force at difference time instances. In particular, the blue curve at $t=7000$ is seen to have a *boundary reaction force along the same direction as the counter flow*. That is in contrast to the Poiseuille flow, where the boundary reaction force is always opposite to the flow direction. For almost all $v_x^C(t)$ prior to reaching the equilibrium state, the flow profile in the vicinity of the boundary is similar to the blue curve in Fig. 6(b). It explains how the *initial counter flow can provide the necessary boundary reaction force for decreasing the net flow*, in accordance to Newton's law. Below we examine in more detail the question of force balance and the necessary condition for a statistical equilibrium (or stationary) state.

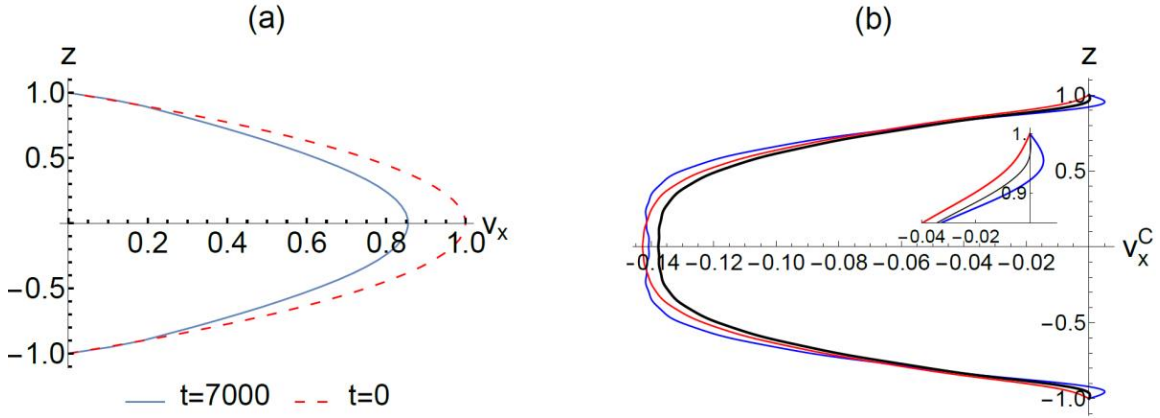

**FIG. 6.** (a) Flow profile at two time instances, where $v_x$ denotes the total velocity. At $t=0$, the profile is parabolic, corresponding to the Poiseuille flow. At $t=7000$, the net flow rate is decreased, even though the profile is still very similar. (b) The counter-flow profile $v_x^C$ at three time instances, obtained by subtracting off the Poiseuille profile from the total flow profile. The inset shows an enlarged view of the velocity profile in the vicinity of the solid boundary. The counter flow profile can be expressed as a superposition of the 1D symmetric modes, not all flowing along the same direction. It is noted that the boundary reaction force is in the negative $x$ direction at $t=7000$ (blue curve). At other two time instances, $t=7560$ (red) and $t=7345$ (black), the boundary reaction force is either along the positive $x$ direction or nearly zero, respectively. The fluctuating boundary profile as a function of time has a direct bearing on the time-averaged frictional reaction force from the solid boundary arising from $v_x^C$. The blue curve is representative of almost all the counter flow profiles prior to $t=7000$.

## E. Force accounting

The NS equation expresses Newton's law in the form $\partial \boldsymbol{v}/\partial t + (\boldsymbol{v}\cdot\nabla)\boldsymbol{v} = \nabla\cdot\overleftrightarrow{\tau}$, where $\overleftrightarrow{\tau}$ denotes fluid's stress tensor. In view of the findings above, an interesting question arises: Is it possible to have a statistical stationary (or equilibrium) state at Re>$Re_c$? A stationary state would imply the vanishing of the inertial force on the left hand side of the NS equation. At $t=0$, the Poiseuille flow is obviously stationary, but in the presence of counter flow at $t=7000$, the vanishing of the inertial force seems very unlikely, even if it is on the statistical basis of taking the time-averaged flow profile. We show below that a detailed force accounting yields a somewhat surprising answer.

The natural starting point is to volume-integrate both sides of the NS equation, and convert the right hand side of the NS equation to a surface integral over the boundaries of the computational domain so that the integrated equation expresses the equality of the total inertial acceleration to the total external forces exerted on the system. The latter comprises the force from the externally applied pressure gradient, plus the frictional reaction force from the solid/liquid interfaces. Since the applied pressure gradient is always given by $-2/\mathrm{Re}$, hence we denote $4LW \times 2 \times \frac{2}{\mathrm{Re}} = F^0$ as the unit of force (along the $x$ direction) in this accounting exercise, i.e., total force from applied pressure gradient=1. By denoting the counter flow velocity profile as $v_x^C$, it is clear that the frictional boundary reaction force $\Delta F_x^{boundary}$ arising from $v_x^C$ must be equal to the inertial force $\Delta F_x^{inertial}$, since pressure force is completely compensated by the boundary reaction force from $u_x^P$, and the total boundary reaction force is the linear addition of that from $v_x^C$ and $u_x^P$. In Fig. 7(a), we plot the inertial force $\Delta F_x^{inertial}$ as a function of time (blue dots), together with the boundary reaction force $\Delta F_x^{boundary}$ caused by $v_x^C$ (red line). Complete agreement is seen, since the two represent the two sides of the same equation.

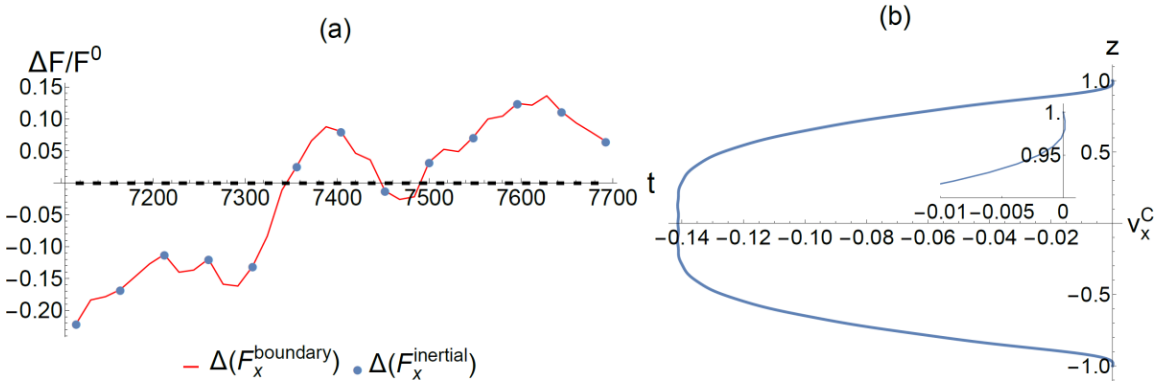

**FIG. 7.** (a) A plot of the normalized inertial force and the boundary reaction force as a function of time. The time period is chosen to be in the near-equilibrium state, i.e., after $t=7000$. The blue dots denote the inertial force, and the red curve denotes the boundary reaction force. Complete agreement is seen, since the two represent the two sides of the same equation. The black dashed line represents the time average of the boundary reaction force. It is very close to zero. (b) The $v_x^C$ profile time-averaged over the same time period as shown in (a). The inset shows an enlarge picture of the profile in the vicinity of the boundary. It is seen that $\left[\frac{\partial v_x^C}{\partial z}\right]_{z=\pm 1} \approx 0$, indicating a near-zero boundary reaction force in the time-averaged sense.

The fluctuating nature of the curve in Fig. 7(a) raises an obvious question: What if one takes a time average over this (near-equilibrium) time period? The result is shown as the flat dashed line in Fig. 7(a), which is slightly negative but very close to zero, i.e., the time-averaged boundary reaction force caused by the counter flow is nearly zero. In Fig. 7(b) we plot a time-averaged profile of $v_x^C$ over the same period. It is seen that at the upper and lower boundaries, $\left[\frac{\partial v_x^C}{\partial z}\right]_{z=\pm 1} \approx 0$, indicating a close-to-zero frictional reaction force from the boundary! This is a somewhat surprising result from our limited data, but consistent with the finding of a statistical equilibrium state that requires the absence of inertial force in the time-averaged sense. This scenario is conceivable because $v_x^C$ comprises different 1D symmetric modes with *both + and − flow directions*, thereby making it possible to have no net boundary reaction force via cancellations. This point has already been foreshadowed in Fig. 6(b), which shows that there can be opposite slopes of $\left[\frac{\partial v_x^C}{\partial z}\right]_{z=\pm 1}$ at different time instances in the vicinity of the boundary, including one that shows $\left[\frac{\partial v_x^C}{\partial z}\right]_{z=\pm 1} \cong 0$ at *t*=7345.

In summary, our analysis yields a physical picture of the equilibrium state beyond Re$_c$ with the following characteristics: (1) fluctuations; (2) co-existence between Poiseuille flow, 1D symmetric modes, and the vortices, with the transient nature of the latter underling the fluctuations; and (3) reduced flow rate from the Poiseuille flow.

### F. Experimental possibilities

Our numerical experiment was limited by the available computational resources. Hence it is not possible to test whether the final result would be altered by considering a very large set of initial states, or by considering many more coupling possibilities via the tensorial term. However, experimental verification of the predictions should be possible. Viscosity can be varied by mixing glycerol and water in varying proportions, for example. Two experiments can be carried out by using the same channel but with different applied pressure across the sample, with compensating viscosity values so as to maintain the same $U_0$ associated with the Poiseuille profile. The value of Re can thus be varied, with one experiment below the critical state and another one above it. In this manner the flow rate of the two experiments can be measured and compared. In another experiment, perhaps more difficult, is to visualize the 2D critical oscillating state, if even in a transient condition.

## V. Conclusion

By obtaining the analytical eigenfunctions of the linearized incompressible NS equations in the channel geometry, we reduced the full NS equation to a set of coupled autonomous ODEs with *t* being the only variable. The critical Reynolds number and the critical wavevector were obtained to five and four significant figures, respectively, together with an explicit configuration of the oscillating critical state. By considering only a limited number of tensorial couplings, and with a small set of initial perturbed HMs, we reveal that beyond the critical state, the net flow rate decreases due to the nonlinear coupling between the 1D symmetric HMs and other HMs with zero momentum. There is a plateau regime for the net flow rate, reduced by ~15% from that in the linear regime, almost independent of the slip length. The 1D symmetric modes are shown to co-exist with the Poiseuille flow and the vortices. However, it remains an open question as to the ultimate net flow rate in the limit of $t \to \infty$, or $\text{Re} \to \infty$, or both. The resolution of this question may require mathematical analysis of the problem, which is beyond the scope of this work.

**Author Contributions** PS initiated and designed the research, XD obtained the analytical 3D eigenfunctions of the incompressible NS equation in the channel geometry, both contributed to the formulation of the channel flow problem and writing of the manuscript.

**Acknowledgement.** PS acknowledges support by the Research Grants Council of Hong Kong Grant No. 16303918. XD thanks Xiaoping Wang for useful discussions.